\documentclass[10pt,twocolumn]{article}

\usepackage[margin=1in]{geometry}  
\usepackage{graphicx}              
\usepackage{amsmath,amssymb}       
\usepackage{cite}                  
\usepackage{listings}              
\usepackage{xcolor}                
\usepackage{booktabs}              
\usepackage{caption}               
\usepackage{hyperref}              
\usepackage{url}

\lstdefinelanguage{JavaScript}{
  keywords={typeof, new, true, false, catch, function, return, null, catch, switch, var, if, in, while, do, else, case, break, const, let, async, await, import, export, from, class, extends, super, this},
  keywordstyle=\color{blue}\bfseries,
  ndkeywords={boolean, throw, import, export, typeof, class, extends},
  ndkeywordstyle=\color{blue}\bfseries,
  identifierstyle=\color{black},
  sensitive=false,
  comment=[l]{//},
  morecomment=[s]{/*}{*/},
  commentstyle=\color{gray}\ttfamily,
  stringstyle=\color{red}\ttfamily,
  morestring=[b]',
  morestring=[b]"
}

\title{Extending ResourceLink: Patterns for Large Dataset Processing in MCP Applications}

\author{
    Scott Frees\\
    Ramapo College of New Jersey \\
    Department of Computer Science \\
    Mahwah, New Jersey \\
    \texttt{sfrees@ramapo.edu}
}

\date{}

\begin{document}

\maketitle

\begin{abstract}

Large language models translate natural language into database queries, yet context window limitations prevent  direct deployment in reporting systems where complete datasets exhaust available tokens. The Model Context Protocol specification defines ResourceLink for referencing external resources, but practical patterns for implementing scalable reporting architectures remain undocumented. This paper presents patterns for building LLM-powered reporting systems that decouple query generation from data retrieval. We introduce a dual-response pattern extending ResourceLink to support both iterative query refinement and out-of-band data access, accompanied by patterns for multi-tenant security and resource lifecycle management. These patterns address fundamental challenges in LLM-driven reporting applications and provide practical guidance for developers building them.

\end{abstract}

\section{Introduction}
\label{sec:intro}

Large language models (LLMs) serve two fundamentally different roles in enterprise data tools - \textit{analysis} and \textit{retrieval}.  For analysis tasks, the LLM ingests and synthesizes data within context window to provide discrete answers to a user; e.g. ``What is the top trending product in terms of sales nation-wide``. In contrast, retrieval tasks use the LLM to generate queries from natural language - where the results are presented to the user or used downstream for data exports and visualizations; e.g. ``Show me the nation-wide sales of our products``.  

This distinction significantly impacts context window utilization. Analysis exhausts tokens proportional to result size, creating computational overhead and latency. However, retrieval tasks need only schema-representative samples or aggregate metadata for query validation. Once the LLM generates a valid query, the host application may execute it independently and render visualizations without LLM involvement, reducing token consumption while maintaining natural language interfaces.

MCP specifications define ResourceLink \cite{mcp2024spec} for referencing rather than embedding data in tool responses, however patterns for reporting tool architectures remain undocumented. Existing literature focuses on conversational analysis workflows where LLMs consume complete datasets, not query generation scenarios. We address this gap by introducing a dual-response pattern combining ResourceLink with sampling validation and metadata for iterative query refinement. We also present multi-tenant server strategies for data isolation and resource discovery mechanisms for dynamic introspection. 

\section{Background}
\label{sec:background}

\subsection{Context Window Budget}
LLM's typically use MCP tools to access data when used in reporting systems, and tool responses place strain context window limitations. The self-attention mechanism's quadratic complexity \cite{dumankeles2022computational} requires O(n²·d) time and memory, creating direct conflict between reporting scale and inference performance. Extended contexts degrade both latency and accuracy.  Liu et. al \cite{liu2023lost} demonstrated LLMs exhibit the ``lost in the middle`` phenomenon, struggling with information in extended sequences. Leng et al. \cite{leng2024longcontext} found only a subset of models maintain accuracy above 64,000 tokens, with substantial latency increases. 

\subsection{Tool Call Context Consumption}
Tool specifications and results consume substantial context budget. When LLMs invoke external tools for reporting data, function parameters, metadata, and returned structured data all occupy context capacity. Gim et al. \cite{gim2024asynchronous} noted synchronous tool-use paradigms exacerbate this - each tool call's complete result must be incorporated before subsequent reasoning. Reporting pipelines invoking multiple sources may find results occupy 70-80\% of context before analysis begins, forcing a choice between comprehensive data and reasoning capacity.  

\subsection{MCP ResourceLink}
MCP specification version 2025-06-18 \cite{mcp2024spec} introduced ResourceLink, enabling servers to reference resources via URI-based handles rather than embedding complete payloads. Resource links provide references without inline transmission. The schema includes URI, description, MIME type, and size, enabling clients to understand characteristics without retrieving full contents. Resource links are ephemeral in nature, they do not represent (necessarily) persistent entities on the server, rather \textit{artifacts} of tools invocations. When reporting tools return ResourceLinks to query results, LLMs receive only handles and metadata - actual results remain unconsumed. Host applications subsequently retrieve full datasets via URI for rendering, decoupling query formation from data transmission.

\subsection{Zero-Shot Accuracy Challenge}
Leading text-to-SQL systems achieve execution accuracy exceeding 80\% on cross-domain benchmarks \cite{liu2025survey}, yet this implies 15-20\% error rates invisible to end users lacking technical expertise to validate outputs. Error rates reduce when models iterate, observing output and refining queries \cite{xia2024r3, askari2024magic, zhang2025dataaware}.

ResourceLink's handle-based architecture suggests LLMs generate correct queries without observing execution samples. Without validating expected schema, row counts, or distributions, LLMs cannot detect errors in joins, aggregations, or filters until rendering. This produces syntactically valid but semantically incorrect queries. This necessitates patterns enabling iterative refinement through result inspection rather than zero-shot generation.

\subsection{User Experience Needs}
Effective LLM-based reporting must support exploratory questioning, iterative refinement \cite{sprague1980framework}, and \textbf{persistent} artifacts reflecting updated data \cite{albo2016visualization}. Business intelligence research shows systems must support goal-oriented workflows with multi-dimensional manipulation \cite{seo2005rank}. Substantive work such as data retrieval, transformation, and rendering occurs in specialized backend services \cite{chen2012business}. This necessitates hybrid context management: resource handles or opaque identifiers reference server-side results, allowing LLMs to reason about structures without materializing contents, while small samples or summary statistics enable direct question-answering. This dual-mode operation (lightweight samples for exploration, handle-based references for artifacts) preserves context for multi-turn dialogues while enabling reports backed by arbitrarily large datasets.

\section{Dual Response Pattern}

The Model Context Protocol specification defines a ResourceLink primitive for referencing external resources within tool responses:

\begin{lstlisting}[caption={ResourceLink Schema as defined in MCP Specifications}, label=lst:javascript]
interface ResourceLink {
    uri: string;            
    name: string;          
    description?: string;  
    mimeType?: string;     
    size?: number;  // bytes       
}

\end{lstlisting}

\subsection{Extending for Dual Response}
We extend ResourceLink with a dual response pattern that addresses the fundamental tension between LLM reasoning requirements and scalable data access. This pattern augments tool responses with two distinct components: (1) preview data suitable for LLM analysis, and (2) a ResourceLink for out-of-band retrieval of complete datasets.

The extended response schema incorporates both immediate analytical samples and persistent resource references. This design enables LLMs to perform immediate analysis on representative samples while preserving access to complete datasets for comprehensive reporting. The preview data flows through the LLM context window, enabling pattern recognition, validation, and direct question answering for queries where samples suffice. The ResourceLink provides a stable identifier for subsequent pagination and retrieval operations that bypass the context window entirely.

\begin{lstlisting}[caption={Expanded schema for dual response tool results}, label=lst:javascript]
interface DualResponseToolResult {
 // Limited results for LLM reasoning
 results: Array<Record<string, any>>;
 // Reference to complete dataset
 resource: ResourceLink;    
 // Query context and constraints
 metadata: QueryMetadata;        
}

interface QueryMetadata {
 // Total records matching query
 total_count: number;          
 // ISO 8601 timestamp
 executed_at: string;             
 // Schema information (optional)
 columns: ColumnDefinition[];    
 // Resource expiration timestamp
 expires_at?: string;           
}


\end{lstlisting}

\begin{figure*}[t]
    \centering
    \includegraphics[width=0.95\textwidth]{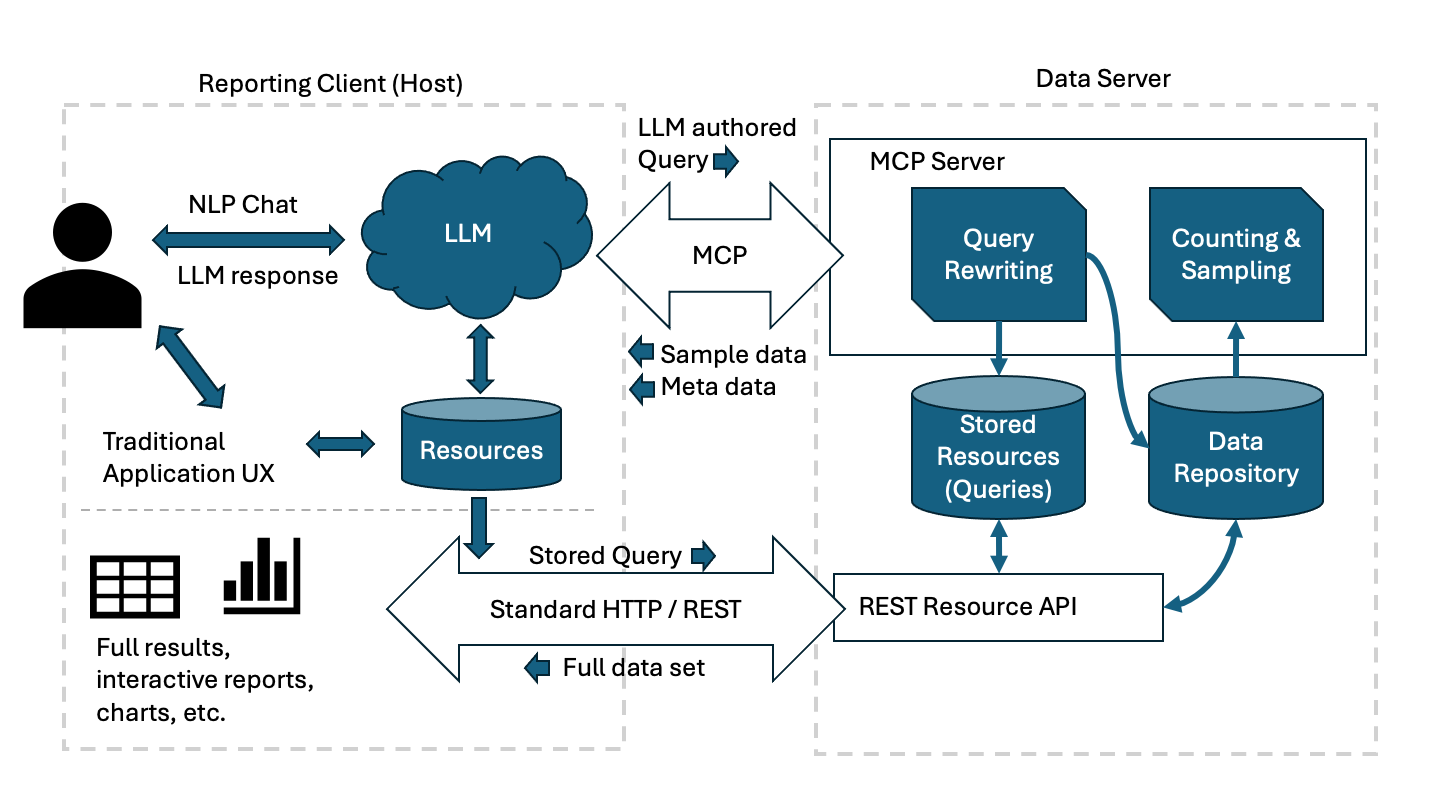}
    \caption{General architecture for dual response pattern integrating LLM tool calls with out-of-band data retrieval. The MCP server returns both preview samples for LLM inference and ResourceLinks for complete dataset access, after altering or augmenting query to provide multi-tenant protection and data sampling. Clients retrieve full data through RESTful endpoints, enabling  reporting without consuming context.}
    \label{fig:yourlabel} 
\end{figure*}

\subsection{Abstraction and Execution}
The pattern deliberately abstracts query semantics to support heterogeneous backend systems; e.g. SQL databases, document stores with aggregation pipelines, graph query languages, or custom analytical engines. The server receives structured query specifications from LLM-generated tool calls and returns both preview samples and resource identifiers regardless of underlying implementation.  Tools prompt the LLM to generate queries in the format the server can interpret.

A critical consideration involves sampling strategies. While random sampling provides statistical representativeness, servers should faithfully execute LLM-specified query constraints including ordering, limiting, and filtering operations. An LLM requesting ORDER BY timestamp DESC LIMIT 10 may be specifically seeking recent records for temporal analysis; substituting random samples would violate the query semantics. We recommend implementing preview generation by applying the complete query specification with an additional limit constraint, ensuring the preview represents what the LLM actually requested rather than an arbitrary sample.  

\subsection{Resource Lifecycle}
Servers may implement resource persistence through multiple strategies: (1) storing complete query results, (2) storing query definitions for re-execution, or (3) hybrid approaches with time-bounded caching. For workloads with frequently-changing data, storing queries enables fresher results on subsequent accesses, albeit with re-execution costs and potential consistency variations.

Developers implementing this pattern must recognize that LLMs generate queries speculatively during iterative refinement processes. As an LLM explores a problem space by testing different aggregations, refining filters, or constructing complex joins, it may invoke query tools dozens of times before producing a final response. Each invocation creates a new resource with an associated ResourceLink, yet the user remains unaware of these intermediate queries. This speculative execution model necessitates aggressive lifecycle management to prevent unbounded resource accumulation.

Resource expiration represents the primary lifecycle mechanism. The \texttt{expires\_at} timestamp in QueryMetadata enables automatic garbage collection of queries. Users may wish to preserve specific queries for reuse in dashboards, reports, or recurring analyses - which is supported by lifecycle endpoints outlined in the next section.

\subsection{Data Limiting}
To prevent context window exhaustion MCP servers should enforce limits on preview responses through query rewriting or post-processing, regardless of LLM-specified parameters. Even when an LLM generates a query with LIMIT 10000, the server should cap data at a reasonable threshold (typically 10-100 records) while accurately reporting \texttt{total\_count} based on the original query.

This dual responsibility requires careful implementation. For SQL backends, servers may wrap queries with additional LIMIT clauses to restrict sample size while executing separate COUNT(*) queries with identical WHERE and JOIN logic to compute \texttt{total\_count}. For aggregation pipelines, servers inject limits as terminal stages while computing counts through parallel execution. For API-backed sources, servers retrieve limited result sets while extracting pagination metadata for totals.

Critically, both preview limiting and count computation must respect multi-tenant isolation filters. The \texttt{total\_count} reflects records accessible to the authenticated user's tenant scope, ensuring preview samples and counts remain consistent with eventual full dataset retrieval through ResourceLink access.

\subsection{Prompt integration}
This pattern integrates naturally with structured output specifications where LLMs return conformant JSON schemas. Tool definitions specify the dual response format as the required return type, constraining LLM behavior through schema validation rather than prompt engineering alone. System prompts should explicitly instruct LLMs on decision logic: analyze preview data for questions answerable from samples; return ResourceLink references for comprehensive reports, exports, or visualizations requiring complete datasets.

\section{Out-of-Band Retrieval}
While MCP tools facilitate query construction and preview generation, complete dataset retrieval occurs through RESTful HTTP endpoints that bypass the LLM context entirely. This architectural separation enables client applications to implement sophisticated reporting interfaces; e.g. paginated tables, interactive visualizations, bulk exports - without consuming context window resources. The MCP server exposes a consistent resource URL prefix (e.g., https://server.com/resources/) that clients discover through server capability negotiation, discussed below.

\subsection{Metadata Fetch}
Client applications retrieve current resource metadata through HTTP GET requests to the resource endpoint:

\begin{lstlisting}[caption={Fetching resource metadata}, label=lst:javascript]
GET /resources/{resourceId}
Authorization: Bearer {token}
\end{lstlisting}

The metadata response provides essential information for client-side rendering and validation, and is the same schema provided by the MCP tool response:

\begin{lstlisting}[caption={Resource metadata response}, label=lst:javascript]
interface QueryMetadata {
    total_count: number;         
    executed_at: string;           
    columns: ColumnDefinition[];   
    expires_at?: string;           
}
\end{lstlisting}

The \texttt{total\_count} field may differ from the original preview response if the underlying data source has changed, enabling clients to detect stale references. The status field allows graceful handling of expired or processing resources before attempting data retrieval.  Servers may update \texttt{expires\_at} timestamps on access to extend lifetimes for frequently-used queries.

\subsection{Data Retrieval}
Full dataset access occurs through HTTP POST requests with pagination and sorting parameters:

\begin{lstlisting}[caption={Retrieving full resource data}, label=lst:javascript]
POST /resources/{resourceId}
Authorization: Bearer {token}
Content-Type: application/json

{
    "offset": 0,
    "limit": 1000,
    "sort": {
        "field": "timestamp",
        "order": "desc"
    }
}

\end{lstlisting}

Servers return structured responses with pagination metadata enabling sequential page retrieval:

\begin{lstlisting}[caption={Full resource data response}, label=lst:javascript]
interface DataResponse {
    total_count: number;
    returned_count: number;
    offset: number;
    data: Array<Record<string, any>>;

    pagination: {
        has_next: boolean;
        has_previous: boolean;
        next_offset: number;
    };
}

\end{lstlisting}

Note that this pattern diverges from the MCP ResourceLink specification which suggests opaque cursor-based pagination where servers return encoded position tokens (e.g., \texttt{next\_cursor: "eyJpZCI6MTIzNH0="}). While cursors prevent inconsistencies when underlying data changes between requests, offset-based pagination offers superior developer ergonomics through (1) direct page access without sequential traversal, (2) stateless server implementation, (3) natural mapping to conventional UI patterns ("showing 1-100 of 5,000") and (4) immediately readable URLs for debugging.

\subsection{Resource Lifecycle Management}
Clients manage resource persistence through HTTP PUT and DELETE operations. The "pinning" operation removes automatic expiration, converting ephemeral queries to persistent artifacts. The effect is the removal of the \texttt{expires\_at} timestamp, allowing indefinite retention until explicit deletion:

\begin{lstlisting}[caption={Saving and deleting a resource}, label=lst:javascript]
PUT /resources/{resourceId}
DELETE /resources/{resourceId}
\end{lstlisting}

This operation typically occurs when users explicitly save queries, incorporate them into dashboards, or schedule recurring reports. Conversely, explicit deletion through HTTP DELETE enables immediate cleanup of unwanted resources without waiting for expiration timers.  The reader can also explore utilizing PATCH for edits where applicable.

\section{Multi-Tenant Access}
Implementing secure multi-tenant query access requires careful data isolation. We identify five distinct patterns with specific trade-offs for MCP server implementations.

\textbf{Pre-filtering and Stage Whitelisting}: Frameworks like MongoDB enable runtime query modification by injecting tenant-specific \texttt{t\$match} stages before user operations. Servers maintain mappings of collections to tenant-identifying fields and automatically prepend isolation filters. This requires whitelisting permitted aggregation stages—excluding \texttt{\$out}, \texttt{\$merge}, and write operations—while handling \texttt{\$lookup} stages to prevent cross-tenant joins.

\textbf{Row-Level Security and Read-Only Connections}: PostgreSQL and similar databases support native row-level security (RLS) policies that automatically filter query results based on session context \cite{liu2025survey}. MCP servers establish connections with tenant-specific credentials, where database policies restrict visibility to authorized rows. Combined with read-only connections, this delegates security enforcement to the database engine. This proves particularly robust though it requires per-tenant connection pooling \cite{liu2025survey}.

\textbf{SQL Abstract Syntax Tree Rewriting}: For systems lacking native RLS, SQL parsing and rewriting provides similar protections. Servers parse LLM-generated SQL into abstract syntax trees, inject tenant predicates into WHERE clauses, and validate table access against whitelists \cite{rizvi2004extending}. This enables fine-grained control but requires sophisticated parsing logic to handle dialect variations and complex subqueries.

\textbf{View-Based Simplification}: Restricting MCP tools to pre-defined database views rather than direct table access naturally encapsulates security logic, denormalizes schemas for LLM comprehension, and mitigate injection attacks. This can simplify both security implementation and LLM prompt engineering.

\textbf{API-Mediated Access}: Exposing RESTful endpoints instead of direct query interfaces maximizes security control but constrains LLM analytical capabilities, limiting users to pre-anticipated access patterns.

\textbf{Authentication and Authorization}:  OAuth 2.0 authenticates MCP tool invocations and authorizes REST requests to resource endpoints. The Bearer token should encode or allow derivation of authorization metadata: tenant identifiers, resource scope, permission levels.  This rich context enables the multi-tenant strategies described.

\section{Discovery}
MCP servers supporting out-of-band resource access advertise capabilities through two complementary mechanisms.  During MCP initialization, servers may declare resource endpoint support.  This minimal advertisement indicates: (1) tools return dual responses with preview data and ResourceLinks, and (2) where REST endpoints for resource access are located. Clients can construct resource URLs by appending ResourceLink URIs to the baseUrl.

\begin{lstlisting}[caption={MCP initialization }, label=lst:javascript]
{
  "protocolVersion": "2025-06-18",

  "capabilities": {
    "tools": {},
    "resources": {
      "resourceLinks": {
        "dualResponse": true,
        "baseUrl": "https://server.example.com/resources"
      }
    }
  }
}
\end{lstlisting}

Following OAuth 2.0's .well-known pattern \cite{hardt2012oauth}, servers may expose complete operational metadata at /.well-known/resource-link-service found at the baseUrl defined in the MCP initialization exchange.

\begin{lstlisting}[caption={REST endpoing specification}, label=lst:javascript]
{
  "methods": {
    "metadata": {
      "method": "GET",
      "path": "/{id}"
    },
    "data": {
      "method": "POST",
      "path": "/{id}",
      "accepts": ["offset", "limit", "sort"]
    },
    "save": {
      "method": "PUT",
      "path": "/{id}"
    },
    "delete": {
      "method": "DELETE",
      "path": "/{id}"
    }
  }
}
\end{lstlisting}

\section{Agent and Tool Prompting}

\textbf{LLM Output Schema Enforcement}: 
Clients use structured output schemas to constrain LLM responses, ensuring resource links from tool calls are returned alongside text responses. Prompts instruct the LLM to: extract  \texttt{uri},  \texttt{name}, and  \texttt{mimeType} from  \texttt{resource\_link} objects; analyze preview data when \texttt{total\_count} $\leq$ \texttt{results.length}; and include resources links when \texttt{total\_count} exceeds preview size or users request artifacts.

\textbf{Derivative Artifacts}: 
Additional MCP tools can create visualizations or dashboards by accepting resource identifiers as inputs and generating new resources. The LLM invokes these sequentially - querying data first, then passing resourceIds to visualization tools.

\textbf{MCP Server Tool Design Patterns}: 
Effective MCP servers implement progressive schema discovery through hierarchical tools: collection/Entity discovery tools with fuzzy search for natural language queries; Property/field schema tools providing detailed metadata with batch retrieval; query tools returning DualResponseToolResult structures; and help tools embedding query patterns. 

\section{Conclusion}
\label{sec:conclusion}

This paper describes a dual-response pattern for MCP ResourceLink implementations, addressing fundamental tensions between LLM context constraints and enterprise reporting requirements. By returning samples for LLM reasoning and resource links for out-of-band data retrieval, we enables query construction while preserving iterative refinement necessary for semantic correctness. Our complementary patterns - multi-tenant isolation strategies, resource lifecycle management, and progressive discovery - establish an architectural framework for production reporting systems where natural language interfaces can drive query generation while result rendering occurs outside the model's context.

Future work centers on community-driven standardization of these patterns. We envision formal specification through MCP enhancement proposals or dedicated RFCs that enable interoperability. Standardized discovery mechanisms, consistent REST endpoint contracts, and shared authentication patterns would accelerate ecosystem development and reduce implementation fragmentation. 

\bibliographystyle{plain}
\bibliography{references}

\end{document}